%Paper: hep-ph/9301246
%From: FTLIPKIN@WEIZMANN.WEIZMANN.AC.IL
%Date: 18 Jan 93   10:08 +02

%macropackage=phyzzx
\magnification=\magstep1
\def\({[}
\def\){]}

\def\rarrow{\rightarrow}

\baselineskip 12pt
\overfullrule 0pt

\bigskip
\centerline{\bf QUARK MODELS AND QUARK PHENOMENOLOGY}

\centerline{\bf Invited Talk at Third Symposium on the History of
Particle Physics}
%\bk
\author{Harry J. Lipkin}
\smallskip
\centerline{Department of Nuclear Physics}
\centerline{\it Weizmann Institute of Science}
\centerline{Rehovot 76100, Israel}

%\bk
\centerline{and}

%\bk
\centerline{School of Physics and Astronomy}
\centerline{Raymond and Beverly Sackler Faculty of Exact Sciences
}
\centerline{\it Tel Aviv University}
\centerline{Tel Aviv, Israel}
%\bk
\centerline{September 18, 1992}

\abstract
\medskip
Overwhelming experimental evidence for quarks as real physical
constituents of hadrons along with the QCD analogs of the Balmer
Formula, Bohr Atom and Schroedinger Equation already existed in 1966.
A model of colored quarks interacting with a one-gluon-exchange potential
explained the systematics of the meson and baryon spectrum and gave
a hadron mass formula in surprising agreement with experiment.
The simple quark model dismissed as heresy and witchcraft by the establishment
predicted quantum numbers of an enormous number of
hadronic states as well as relations between masses, reaction cross sections
and electromagnetic properties, all unexplained by other approaches.
Further developments leading to QCD included confinement in the large
$N_c$ limit, duality, dual resonance and string models,
high energy scattering systematics, unified treatment of mesons and
baryons, no exotics and no free quarks.

\medskip
\centerline{\bf I. PROLOGUE - HOW TO THINK ABOUT QUARKS}

\medskip
\centerline{\bf 1.1  Dedication - Implications of BCS for Quarks}
\medskip

I begin with a tribute to a great physicist who taught me how to think about
quarks and physics in general, John Bardeen. A few sentences from John could
often teach you more and give more deep insight than ten hours of
lectures from almost anyone else.
In 1966 when I began to take quarks seriously I was unknowingly thinking
about them in the language I had learned from John during two years at the
University of Illinois, as quasiparticle degrees of freedom
describing the low-lying elementary excitations of hadronic matter.
Unfortunately I did not realize how much my own thinking had been
influenced by John Bardeen until he was gone. I dedicate this paper to his
memory.

Were quarks real? Quarks as real as Cooper pairs would be enough.
Quarks leading to anything remotely approaching the exciting
physics of BCS would be more than enough. John always
emphasized that Cooper pairs were not bosons, and that superconductivity
was not Bose condensation.
The physics was all in the difference between Cooper pairs and bosons. I
was not disturbed when quarks did not behave according to the
establishment criteria for particles. The physics might all be in the
difference between quarks and normal particles. One had to explore the
physics and see where the quark model led.

The arguments of the BCS
critics that the theory was not gauge invariant did not disturb John; he
knew where the right physics was. Similarly the arguments criticizing
quarks as non-relativistic did not disturb me. The model had the
right physics. It already in 1966 described so much experimental
data not understood by any other model that it had to
have the right physics. The formalism would come later and the
basis of QCD was already published in 1966\REF{\Nambu}{Y. Nambu,
in Preludes in Theoretical Physics, edited by
A. de Shalit, H. Feshbach and L. Van Hove, (North-Holland Publishing Company,
Amsterdam, 1966), p. 133}$[{\Nambu}]$. A model of colored quarks
interacting with colored gauge bosons in the manner described by a
non-Abelian gauge theory had so much of the right physics\REF{\SakhZel}{
Ya. B. Zeldovich and A.D. Sakharov, Yad. Fiz 4 (1966)395; Sov. J. Nucl. Phys.
4 (1967) 283}$[{\SakhZel}]$ that it had to lead somewhere.
But there are none so blind as those who don't want to see.

\medskip
\centerline{\bf 1.2 A Historical Perspective}
\medskip

The history of this period can be characterized by repetition at
successive levels of the conflict between ``Grand Unification" and
``Compositeness" approaches to the structure of
matter. Each stage began with the belief that the fundamental
constituents of matter or ``elements" were known. The experimental
discoveries of too many elements led on the one hand to attempts to
unify the elements while still considering them as elementary, and on the other
to build them out of a smaller number of fundamental building blocks. In 1950
the nucleon and pion were considered the fundamental constituents of
hadronic matter. Evidence for composite structure was
resisted by the establishment who sought to unify the large number of new
``elementary" particles with concepts like nuclear democracy or
higher symmetry, in which all particles were equally elementary. Today we
have come full circle back to square one at a deeper level. All matter is
constructed from quarks and leptons. The explanations of the large number of
elementary objects using grand unification or compositeness have moved from the
nucleon-pion level to the quark-lepton level.

The quark model developed very differently in the Eastern and Western
Hemispheres. In the East the model was taken seriously from the
beginning and supported by top establishment figures like Bogoliubov,
Sakharov, Zeldovich, Gribov, Thirring, Morpurgo and Dalitz.
The Western approach was stated explicitly by M. L. Goldberger in
introducing a colloquium speaker at Princeton in 1967. ``A boy was
standing on a street corner snapping his fingers and claiming that it
kept the elephants away. When told that there had been no elephants
around for  many years, his response was `You see! It works!'. And
now our speaker will talk about the quark model."

The approach of Galileo of studying nature by experiments led Eastern physics
to the conclusion ``The quark model works, and we do not understand it.
Therefore it is interesting." Western theorists who seemed to have forgotten
Galileo concluded ``The quark model works, but it contradicts the
established dogma. Therefore it is heresy and witchcraft."

A true perspective requires distinguishing between dogma, phenomenology
that contradicts established dogma but works, and phenomenology which
contradicts established dogma but does not really work and is nonsense.
The quark model really worked and pointed the way toward future new ideas
and a new and better understanding of the structure of matter. Two
interesting examples in today's physics are high $T_c$ superconductivity
and cold fusion. Both surprised everybody when they were first announced.
But high $T_c$ really works and demands further investigation for a
better understanding. Cold fusion is nonsense and does not work.

Israeli particle physics was at the crossroads between East and West with roots
in Moscow Leningrad and London. In 1967-68 when Goldberger referred to the
quark model as witchcraft, a group of young junior faculty and postdocs
named Rubinstein, Veneziano, Virasoro, Horn, Harari and Rosner who had come to
Israel after spending time in the West were putting
the new quark model ideas together with accepted S-matrix Reggeism. Thus
began a new era in particle physics then called duality which laid the
foundations for what is now called string theory
\REF{\Venez}{G. Veneziano, Phys. Reports 9C (1974) 199}
$\({\Venez}\)$.

\medskip
\centerline{\bf 1.3 Weak and Strong SU(3) - Constituent and Current Quarks}
\medskip

Murray Gell-Mann pinpointed an important ingredient in understanding quarks:
the difference between ``weak" and ``strong" SU(3) flavor algebras which
led to constituent and current quarks.
Two independent breakthroughs were based on quark-like degrees of freedom.
That QCD had the right physics to describe strong interaction dynamics was
already clear in 1966, with constituent quarks interpreted as quasiparticle
degrees of freedom describing elementary excitations. But current quarks
then only provided a mathematical basis for current algebra and were not seen
as
real physical point-like objects until the quark-parton description of SLAC
experiments. The relation between constituent and current quarks is expected
to come somehow out of QCD, but may
well be as difficult as getting BCS out of the Lagrangian of QED.

\medskip
\centerline{\bf II. SOME PREHISTORY}
\centerline{\bf 2.1 Flavor Symmetry and  Composite Models}

An early composite model of hadrons was the Fermi-Yang model of a pion as a
bound nucleon-antinucleon pair. Its generalization by Sakata to include
strange particles and a flavor symmetry generalized from isospin SU(2) to
SU(3) was soon seen to be in conflict with experiment\REF{\LevSal}{
C. A. Levinson, H. J. Lipkin, S. Meshkov, A. Salam and R. Munir,
Physics Letters 1, 44 (1962)}$[{\LevSal}]$.

The ``Eightfold Way" of Gell-Mann and Ne'eman introduced an SU(3) flavor
symmetry and a hadron classification from two different points of view.
Gell-Mann's ``weak SU(3)" began with the properties of the electroweak
currents; Ne'eman's ``strong SU(3)" with a gauge
theory of strong interactions. Both used octet classifications for baryons
and mesons with no theoretical explanation for the octet baryon classification
nor any physical interpretation for the fundamental triplet.
Goldberg and Ne'eman
\REF{\GOLDNEEM}{H. Goldberg and Y. Ne'eman, Nuovo Cimento 27 (1963) 1}
$[{\GOLDNEEM}]$ extended SU(3) to U(3) and included baryon number
in a formulation constructing the baryon octet
from three fundamental triplets carrying baryon number 1/3.
Ne'eman also suggested that SU(3) was an exact symmetry of strong
interactions broken by an additional ``fifth interaction"\REF{\YUVAL}{
Yuval Ne'eman, Phys. Rev. 134 (1964) B1355}$[{\YUVAL}]$.
But the fundamental triplets of U(3) were presented only as an algebraic device
and not as physical particles.

The ``weak" and ``strong" approaches to flavor symmetry are parts of two very
different lines of development of electroweak and strong interaction physics
over the past forty years.
Electroweak physics is characterized by the ``standard model syndrome",
with most experiments either testing a standard model or
looking for new physics beyond it.
In 1945 the standard model for electroweak physics was the Quantum
Electrodynamics in Heitler's book and the Fermi theory of beta decay.
Crises when the standard model appeared to be wrong were resolved by either
revealing wrong experiments or finding new concepts like parity nonconservation
easily fit into the existing framework. The first indications of ``physics
beyond this standard model" arose in infinities in QED calculations and the
Lamb shift experiment and in disagreements between measured beta ray spectra
and Fermi theory. The QED difficulties were solved by the new formulation of
Feynman, Schwinger and Tomonaga. The difficulties with beta ray spectra went
away after better experiments confirmed the Fermi theory. The development
through various similar crises to modern electroweak theory was
straightforward.

Hadron physics developed very differently with no sensible ``standard
model" until QCD. Today's picture of QCD proton structure bears no resemblance
to accepted models of the 1940's, 50's and 60's. The particle theory
establishment clung to old dogma and refused to accept new ideas until forced
by experimental data. Concepts now generally accepted like spontaneously
broken symmetries, chiral symmetry, the unitary symmetry now called
flavor-SU(3), quarks, and the the color degree of freedom were ridiculed by
the reactionary establishment as they were dragged kicking
and screaming along the path that eventually led to QCD.

At the 1960 Rochester Conference I mentioned to Nambu that I had heard from
John Bardeen in Urbana about his very interesting application of ideas from
superconductivity to particle physics. Nambu said I was the only person at the
conference who had expressed any interest in this work. At the 1962 Rochester
conference in Geneva, the prediction that a particle later called the
$ \Omega^- $ should exist, already proposed in a paper by Glashow and Sakurai,
was not considered important enough to be mentioned in any invited or
contributed talk.
It was mentioned in a comment from the floor by Gell-Mann.
The paper proposing the existence of quarks was
accepted by Physics Letters only because it had Gell-Mann's name on it. The
editor said ``The paper looks crazy, but if I accept it and it is nonsense,
everyone will blame Gell-Mann and not Physics letters. If I reject it and it
turns out to be right, I will be ridiculed."

Today we accept Ne'eman's proposal of a non-Abelian gauge theory with exact
flavor symmetry for strong interactions and flavor symmetry breaking by
a completely different interaction. But the basic degrees of freedom are
completely different. The fundamental fermions and gauge bosons are not
Ne'eman's baryon and vector meson octets but colored quarks and gluons, with
more than three flavors and an additional color degree of freedom.
\medskip

\centerline{\bf 2.2 $\bar p p$ Annihilation - First Evidence for Quarks}

Annihilation experiments\REF{\Gerson}{W.H.Barkas et al Phys. Rev. 105
(1957) 1037} $[{\Gerson}]$ performed shortly after the antiproton
discovery gave results disagreeing with conventional model predictions.
A pion multiplicity of $5.3 \pm 0.4$ was found, much greater than the
2 or 3 predicted by statistical models, while $e^+ - e^-$ pairs were not seen
at
the level predicted by QED from one-photon annihilation of a pointlike
$\bar p p$ pair. Pions as quanta of a boson field could be created only after
the annihilation of the positive and negative baryon number present
in the initial state. No one considered the simple but
unacceptably heretical explanation that both mesons and baryons were composite
objects made of the same constituents which carried baryon number,
rather than being elementary and completely different like photons and
electrons, that no annihilation
of baryon number was needed and that constituents with opposite
baryon number simply rearranged to form ``positronium-like" states with
a multiplicity related to the number of constituents originally present.
Shortly after the quark proposal, such a model showed that a
rearrangement of the three quarks and three antiquarks in the
proton and antiproton into three mesons\REF{\RubStern}{
H.R. Rubinstein and H. Stern, Phys. Lett. 21 (1966) 447}
$[{\RubStern}]$ gave the observed pion multiplicity.
A simple ``back-of-the-envelope" calculation for pions produced from
three s-wave $q \bar q$ pairs with the standard 3:1 statistical factor
favoring the spin-triplet $\rho$ which decays into two pions gives
$3\cdot (3/4)\cdot 2 + 3 \cdot (1/4) = (21/4) = 5.25$.

This quark-rearrangement model was ridiculed as nonsense when proposed
$[{\RubStern}]$ in 1966. The establishment prejudice against quarks even
created serious difficulties for obtaining
appointments and promotions for young people in our group. Deans and
committees were influenced by pejorative comments in letters from
well-known physicists about people who rush into print with such garbage.
\medskip

\centerline{\bf 2.3 Group Theory}
\centerline{\bf From Physics Without Groups to Groups Without Physics}

Until the discovery of the $\Omega^-$ the particle physicists believed
that group theory was useless for high energy physics, thought of isospin as
rotations in some three-dimensional space and knew nothing about unitary
groups. They therefore tried rotations in 4, 5, 6, 7 and 8
dimensions with fancy names like global symmetry, cosmic symmetry, etc. before
finding that the natural symmetry group to include the $SU(2) \times U(1)$ of
isospin and strangeness was $SU(3)$. Perhaps they called it the ``Eightfold
Way" because it took them eight years (1953-61) to find it.

Soon afterwards the pendulum swung and a
flood of papers tried to include flavor SU(3) and space-time in a
larger group and produced a number of fancy no-go theorems.
I noted immediately\REF{\Lipgroup}{Harry J. Lipkin, Phys. Lett. 14 (1965) 336
and in ``Methods and Problems of Theoretical Physics", Proceedings of the
Conference, Birmingham, July 1967, Edited by J. E. Bowcock, North-Holland
Publishing Company, Amsterdam (1970) p. 381} $[{\Lipgroup}]$ that the physics
underlying these fancy groups was completely crazy. No sensible
interaction could be invariant under transformations generated by operators
acting nontrivially both in space-time and in an internal symmetry space.
Translation invariance implies that a pion-nucleon scattering experiment
at SLAC gives the same results when moved to Fermilab. Isospin invariance
implies $\sigma(\pi^- p) = \sigma(\pi^+ n)$. But invariance under
transformations acting in space-time like a translation and also transforming
nontrivially under isospin can move a pion beam from a SLAC experiment to
Fermilab, while leaving the nucleon target at SLAC. Any dynamics invariant
under such transformations must obviously have no interactions, no bound states
and a continuous mass spectrum. However, no one paid attention to this kind of
``low-brow phenomenology" and fancy theorems were published showing that
nonsense is nonsense.
\medskip

\centerline{\bf III. STATIC HADRON PROPERTIES IN THE QUARK MODEL}

The significance of quark model predictions has been confused by model
builders who produce an apparently large number of predictions from a specific
model without noting that only two or three depend on the model and the rest
all follow from model-independent symmetries like angular momentum, isospin
and SU(3). They get excellent but meaningless $\chi^2$ fits to data.
We avoid the pitfall by considering only those quark model predictions
not easily obtained in other ways, and in particular relations between
mesons and baryons and the determination of the values of parameters
which are left free in SU(3).

\centerline{\bf 3.1 The Very Early Successes}

The difference between the quark structures of the meson and baryon octets
immediately explained striking regularities in the low-lying hadron spectrum
not explained by SU(3); e.g. the baryon octets and decuplets and
meson nonets without the ninth baryon suggested by
some SU(3) models and no meson decuplets and the spin-parity quantum numbers
$J^P=0^-,\ 1^-, 1/2^+,\ 3/2^+$.
Introducing U(3) rather than SU(3) and breaking SU(3) at the
quark level by setting $m_s > m_u$ immediately gave the experimentally
observed mass inequalities
$$ M_{\Xi} > M_{\Sigma} \approx M_{\Lambda}>  M_N; ~ ~ ~ ~
M_{\eta} > M_{K^+} \approx M_{K^-}>  M_{\pi}      \eqno (3.1a) $$
instead of the bad baryon mass inequality
following from using the same structure for baryon and meson octets.
$$ M_{\Lambda}>  M_N \approx M_{\Xi} > M_{\Sigma} \eqno (3.1b) $$
These regularities still did not influence the establishment to take
quarks seriously. Many open questions remained; e.g.
the reason for the decuplet classification for the spin-3/2 baryons, rather
than octet or singlet, the reason for the $\Lambda-\Sigma$ mass difference
and whether the next excited states were orbital excitations or states with
additional $\bar q q$ pairs,
\medskip
\centerline{\bf 3.2 The Relevant Degrees of Freedom}

Thirty years of experimental hadron spectroscopy have failed to produce any
evidence for excitations of any of the additional degrees of freedom proposed
for theoretical reasons; e.g. bags, strings, meson clouds, gluons, and a
sea of $\bar q q$ pairs including strange quarks. All observed
hadronic states are described as excitations of the spins and relative
co-ordinates of the constituent quarks in the $\bar q q$ and $3q$ systems.
There is no evidence for excitations
describable as relative motion between the center-of-mass of the valence
quarks and other constitutents like a bag, cloud or sea. Although the
constitutent quark is not believed to be an elementary point-like object
but rather a more complicated object with internal structure, there is
so far no experimental evidence for low-lying excitations of this
structure; i.e. no evidence for ``excited constituent quarks."
Many model builders have attempted to introduce such additional degrees
of freedom, either to satisfy theoretical prejudices or to obtain a
``better fit" than the simple constituent quark model to certain experimental
data. Any advantages claimed by these models must be
scrutinized carefully before acceptance and the absence of any observed
low-lying excitations of such degrees of freedom must be explained.

\medskip
\centerline{\bf 3.3 SU(6) and the Symmetric Quark Model}

The great breakthrough in baryon spectroscopy was the application of
SU(6) symmetry
\REF{\Feza}{F. G\"ursey and L. A. Radicati, Phys. Rev. Lett. 13 (1964) 173
}
$[{\Feza}]$ with the unreasonable assumption that spin 1/2 quarks
obeyed Bose statistics.  The contradiction was avoided
by the introduction of parastatistics\REF{\OWG}{O. W. Greenberg. Phys.
Rev. Lett. 13 (1964) 598} $[{\OWG}]$
or an additional internal degree of freedom
\REF{\Tavkhel}{A. Tavkhelhelidze, in High Energy Physics and Elementary
Particles, IAEA, Vienna (1965) pp. 753 and 763}$[{\Nambu},{\Tavkhel}]$ later
called color. Great progress was made in understanding the baryon spectrum
without a fundamental understanding of statistics by the phenomenological
``symmetric quark model"\REF{\FRT}{P. Federman, H.R. Rubinstein and I. Talmi,
Phys. Lett. 22 (1966) 203} $[{\OWG},{\FRT}]$ which
classified the hadron spectrum according to the group $SU(6)\times O(3)$.
It described all baryons as three quark states with wave functions
satisfying Bose statistics and having orbital and radial excitations with
quantum numbers qualitatively described by a harmonic oscillator shell
model\REF{\RHD}{R. H. Dalitz, in:
Proceedings of the 1967 Irvine Conference on Pion-Nucleon Scattering
edited by Gordon L. Shaw and David Y. Wong, John Wiley and Sons,
New York (1969) p. 187, and
in Proceedings of the XIIIth International
Conference on High Energy Physics, Berkeley 1966. University of
California Press (1967) p. 215}
\REF{\KARLOB}{G. Karl and E. Obryk, {Nucl. Phys.} {B8} (1968) 609}
$[{\OWG},{\RHD},{\KARLOB}]$.
An enormous number of baryon resonances fit exactly into this
simple potential model
beginning with the SU(6) 56 classification of the
lowest baryons into a spin 1/2 flavor octet and a spin 3/2 decuplet,
the first excited configuration being a $70$ of SU(6) with L=1 and the
second being an L=2 56. But the overwhelming evidence repeatedly
presented by Dalitz et al for this model was consistently$[{\RHD}]$
dismissed by the establishment.

The successful SU(6) prediction of -3/2 for the ratio of the proton and
neutron magnetic moments was again striking evidence for compositeness,
since only a composite model gave a simple ratio for $total$ moments.
In other approaches adding Dirac and anomalous moments was like
adding apples and oranges. The anomalous moment was a function of the
strong interaction coupling constant; the Dirac moment was not.
Meson magnetic moments were not measured directly, but the radiative
magnetic dipole transition $\omega \rightarrow \pi \gamma$ is described
by the same quark magnetic operators appearing in the proton moment. The
successful prediction relating this transition to the proton magnetic
moment\REF{\BecMorp}{C. Becchi and G. Morpurgo, Phys. Rev. 140 (1965)
B687}
$[{\BecMorp}]$ again confirmed that mesons and baryons were made of the
same quarks.

The scale of the nucleon magnetic moments caused
confusion since quark magnetic moments were expected to have the
scale of the quark mass rather than the hadron mass, while detailed
relativistic calculations of hadron properties by the Soviet group
$[{\Tavkhel}]$ gave hadron moments at the right scale.
This was resolved
\REF{\LipTavk}{H. J. Lipkin and A. Tavkhelidze, Phys. Lett. 17 (1965)
331}
$[{\LipTavk}]$ by noting that the effective mass appearing in the
magnetic moment of a bound Dirac particle depends upon the Lorentz
structure of the potential and its scale is set by the particle energy,
not its mass, for a world scalar potential. The relativistic calculations
$[{\Tavkhel}]$ effectively assumed a world scalar potential.

\medskip

\centerline{\bf 3.4 The Pre-History of QCD}

Andrei Sakharov was a pioneer in hadron physics who
took quarks seriously already in 1966.
He asked ``Why are the $\Lambda$ and $\Sigma$ masses
different? They are made of the same quarks!"\REF{\Sakhaut}{Andrei
D. Sakharov, Memoirs, Alfred A. Knopf, New York (1990) p. 261}$[{\Sakhaut}]$.
His answer that the difference arose from
a flavor-dependent hyperfine interaction led to relations
between meson and baryon masses in surprising agreement with
experiment$[{\SakhZel}]$.
Sakharov and Zeldovich $anticipated$ QCD
by assuming a quark model for hadrons with a flavor dependent linear mass
term and a two-body interaction
whose flavor dependence was all in a hyperfine interaction
$$v_{ij} = v^o_{ij} +
{{\vec{\sigma}_i\cdot\vec{\sigma}_j}}v^{hyp}_{ij} \eqno(WW3.2)
$$
where $v^o_{ij}$ is independent of spin and flavor,
$\vec{\sigma}_i$ is a quark spin operator and $v^{hyp}_{ij}$ is
a hyperfine interaction with different strengths
but the same flavor dependence
for $qq$ and $\bar q q$ interactions.
They obtained two relations
between meson and baryon masses in surprising
agreement with experiment
\REF{\Gorky}{A. D. Sakharov, Pisma JETP 21 (1975) 554; JETP 78
(1980) 2113 and 79 (1980) 350}
$[{\SakhZel,\Gorky}]$,

The mass difference between $s$ and $u$ quarks calculated
in two ways from the linear term in meson and baryon masses showed
that it costs exactly the same energy to replace a nonstrange quark by a
strange quark in mesons and baryons, when the contribution from the
hyperfine interaction is removed.
$$
(m_s-m_u)_{Bar}
=M_\Lambda-M_N=177\,{\rm MeV}
\eqno(WW3.3a)
$$
$$
(m_s-m_u)_{Mes} =
{{3(M_{K^{\scriptstyle *}}-M_\rho )
+M_K-M_\pi)}\over 4} =180\,{\rm MeV}
\eqno(WW3.3b)
$$
where the subscripts $u$, $d$ and $s$ refer to quark flavors.
The flavor dependence of the hyperfine splittings calculated
in two ways from meson and baryon masses gave the result
 $$ 1.53 = {{M_\Delta - M_N}\over{M_{\Sigma^*} - M_\Sigma}}
= \left({{v^{hyp}_{ud}}\over{v^{hyp}_{us}}}\right)_{Bar}
= \left({{v^{hyp}_{ud}}\over{v^{hyp}_{us}}}\right)_{Mes}  =
{{M_\rho - M_\pi}\over{M_{K^*}-M_K}}= 1.61 \eqno(WW3.4) $$

This striking evidence that mesons and baryons are made of the same
quarks and described by a universal linear mass formula with spin corrections
in remarkable agreement with experiment was overlooked for amusing reasons
\REF{\HJLCG}{Harry J. Lipkin, in
Gauge Theories, Massive Neutrinos and Proton Decay
(Proceedings of Orbis Scientiae 18th Annual Meeting, Coral Gables,
Florida, 1981) edited by Behram Kursonoglu and Arnold Perlmutter (Plenum,
N.Y., 1981), p. 359}
\REF{\Postcard}{
A. D. Sakharov, private communication; Harry J. Lipkin, Annals of the
New York Academy of Sciences 452 (1985) 79,
and London Times Higher Education Supplement, January 20 (1984), p.17}
$[{\HJLCG},{\Postcard}]$ and rediscovered only in 1978
\REF{\HJLADS}{Harry J. Lipkin, Phys. Lett. {\bf B74} (1978) 399}
$[{\HJLADS}]$.
In that same year 1966 Nambu derived just such a universal linear mass formula
for mesons and baryons from a model in which colored quarks were bound into
color singlet hadrons by an interaction generated by coupling the quarks to
a non-abelian SU(3) color gauge field, and spin effects were
neglected
$[{\Nambu}]$.

The Nobel Prize for QCD might have been awarded to Sakharov,
Zeldovich and Nambu. They had it all in 1966. The Balmer formula, the Bohr
atom and the Schroedinger equation of Strong Interactions. All subsequent
developments leading to QCD were just mathematics and public relations, with
no new physics.
But the particle physics establishment refused to recognize the beginnings of
new physics and had to wait until new fancy names like chromodynamics,
color, confinement, etc. were invented together with a massive public
relations campaign. Then they claimed that they had discovered it all.

\medskip

\centerline{\bf 3.5 Color, Confinement and Large N}

The color degree of freedom solved the quark-statistics problem for
baryons and also provided answers to several puzzles previously
unanswered. The observed hadron spectrum indicated that both
$qq$ and $\bar q q$ interactions were attractive in all
possible states of spin and parity. An antiquark should be attracted
by the three quarks in a baryon to make a $3q \bar q$ bound state.
But there were no bound states with ``exotic" quantum numbers that could
not be made from the $q \bar q$ or $3q$ configurations.
There was also the meson-baryon puzzle - why $qq$ and $\bar q q$
systems are bound but different. No simple meson-exchange model gave these
properties.

In 1967 I noted that quarks would be confined in the limit where the
number of colors was large, now called the large $N_c$ limit
\REF{\LipHouch}{Harry J. Lipkin, in:
"Physique Nucleaire, Les Houches 1968,"
edited by C. de Witt and V. Gillet, Gordon and Breach, New York (1969).
p. 585}
$[{\LipHouch}]$.
$\bar q q$ pairs were bound into mesons, the
meson-meson interaction went to zero, the hadron spectrum was simply
systems of non-interacting mesons and free quarks would not be observed.
At that time any heretical paper of this type would never be accepted by a
reputable refereed journal; I therefore put it into lecture notes.
In 1972 I looked at saturation in toy models of nuclei and noted
that a nucleon-nucleon isospin-exchange interaction produced by $\rho$ exchange
would bind only the deuteron and the isoscalar $N \bar N$ system, and that no
higher mass bound states would exist. This led naturally to replacing
isospin SU(2) by color SU(3) and a model with colored quarks interacting
with a color-exchange potential to give the first explanation of the absence
of exotics and the observed meson-baryon systematics\REF{\LipkTriEx}{H.J.
Lipkin, Phys. Lett. 45B (1973) 267} $[{\LipkTriEx}]$ as well as the relation
between $qq$ and $\bar q q$ potentials later used in all potential
models treating both mesons and baryons.

I was very excited to have found a simple explanation of so much hadron physics
for which there was no other explanation, and wrote letters from Israel to
several friends including Dick Feynman and Viki Weisskopf. Feynman never
answered, but Viki wrote that it was all very interesting but theorists would
not like it because it was not renormalizable. This did not bother me as it
rather reminded me of the criticisms of BCS as not being gauge invariant.
Thinking along the lines of BCS I was sure that I had found interesting physics
and that the correct formalism would come later. In fact the discovery of
asymptotic freedom came at the same time and it is interesting to compare the
situation in the summers of 1972 and 1973. In his summary talk at
the 1972 Rochester Conference at Fermilab, Gell-Mann noted that the color
degree of freedom was established from electroweak data, that strong
interactions were still unknown and would probably arise from exchanges of
vector gluons. But there was no suggestion that color played any role in strong
interactions. At the SLAC summer school in 1973 I was invited to talk about my
work on ``Quarks and colored glue" and Gross, Politzer and Wilczek were talking
about the great breakthrough of asymptotic freedom.

Someone called my attention to
Nambu's old paper $[{\Nambu}]$, the details of which I had forgotten, which had
worked out the SU(3) algebra of this interaction, but not investigated the
spatial dependence or the implications for exotics. In contrast with the
behavior of some of my peers, I immediately rewrote the paper giving Nambu full
credit for the work I had independently rediscovered before submitting the
paper
for publication.

It is rather painful to note the disparaging and untrue criticism of my paper
\REF{\FGL}{H. Fritzsch, M. Gell-Mann and H. Leutwyler, Phys. Lett. 47B (1973)
365} $[{\FGL}]$ : ``Recently this
point has been given publicity by Lipkin $[{\LipkTriEx}]$, who treats, however,
a Han-Nambu picture ..... We have rejected such a picture."
Murray Gell-Mann is a great physicist whose work and ideas
have had a tremendous impact on the work and thinking of practically everyone
attending this history conference including myself. But the general consensus
of those active in the field in 1973 is
that there was nothing new nor original in this paper$[{\FGL}]$.
My paper $[{\LipkTriEx}]$, treats only strong
interactions, ignores electromagnetism and the possibility of integrally
charged quarks and has nothing to do with Han-Nambu.
This irrelevant red herring is discussed below.
Their criticism$[{\FGL}]$ is irrelevant nonsense.

\medskip

\centerline{\bf IV. QUARK MODEL PREDICTIONS FOR HADRON REACTIONS}

\medskip

Further evidence for a quark structure of hadrons was found in
the additive quark model for hadron reactions, the so-called ideal mixing
pattern of vector and tensor mesons,
a mysterious topological quark diagram selection rule now called
OZI\REF{\Okubo}{S. Okubo, Phys. Lett {5} (1963) 1975 and
Phys. Rev. D16 (1977)  2336}
\REF{\Zweig} {G. Zweig, CERN Report No. 8419/TH412 unpublished, 1964;
and in Symmetries in Elementary
Particle Physics (Academic Press, New York, 1965) p. 192}
\REF{\Iizuka} {J. Iizuka, Prog. Theor. Phys. Suppl. 37-38 (1966) 21}
$\({\Okubo,\Zweig,\Iizuka}\)$
and peculiar systematics in the
energy behavior of certain hadron total cross sections.

\medskip

\centerline{\bf 4.1 The Additive Quark Model, Duality and Dual Resonance
Models}

The simple additive quark model (AQM) of Levin and
Frankfurt\REF{\LF}{E. M. Levin and L. L. Frankfurt, Zh. Eksperim. i.
Theor. Fiz.-Pis'ma Redakt {2} (1965) 105; JETP Letters {2} (1965) 65}
$\({\LF}\)$ explained the ratio of 3/2 between nucleon-nucleon and
meson-nucleon
scattering and again showed mesons and baryons to be made of the same quarks.
Further refinements included flavor dependence of the scattering amplitudes at
the quark level\REF{\LS}{H. J. Lipkin and F. Scheck, Phys. Rev. Letters 16
(1966) 71}$\({\LS}\)$. That the total cross sections in channels now
called exotic do not have the sharply decreasing behavior found in other
channels, was be described in the AQM by attributing all the energy decrease to
$\bar q q$ annihilation amplitudes
\REF{\HJLANN}{H. J. Lipkin, Phys. Rev. Letters 16 (1966) 1015}
$\({\HJLANN}\)$. The AQM was combined with a Regge picture attributing this
energy behavior to exchange degeneracy of Regge trajectories by using the AQM
to relate the couplings of hadrons to exchange-degenerate Regge
trajectories\REF{\JENSEN}{H. J. Lipkin, Z. Physik 202 (1967) 414}
\REF{\HJLDD}{Harry J. Lipkin, {Nucl. Phys.} {B9} (1969) 349}
$\({\JENSEN,\HJLDD}\)$.
The universality of
additive quark couplings to mesons and baryons arose again and again in
different contexts in these descriptions.

An S-matrix Regge approach beginning with finite-energy sum rules then led to
duality with the same states appearing both as s-channel resonances and
t-channel exchanges and then to dual resonance models beginning with the
Veneziano model$\({\Venez}\)$.
Although this was not directly related to the quark model, it soon appeared
that
introducing the quark-model constraints on Reggeon couplings provided a
powerful input with predictive power. Thus for example the absence of
exotics both as resonances and t-channel exchanges led to the OZI rule,
while the exchange degeneracy and the dominance of the energy-dependent
part of the cross section by $\bar q q$ annihilation led naturally
to duality diagrams\REF{\HHDD}{H. Harari, {Phys. Rev. Lett.} {22}  (1969) 562}
\REF{\JLRDD}{J.L. Rosner, {Phys. Rev. Lett.} {22}  (1969) 689}
$[{\HHDD},{\JLRDD},{\HJLDD}]$. The energy independent part of the cross
section,
later found to be slowly rising, was seen to be related to diffraction,
described by Pomeron exchange, with a coupling given by the
Levin-Frankfurt quark-counting recipe.
\medskip

\centerline{\bf 4.2 Neutral Meson Mixing, OZI and the November Revolution}

The first use of the additive quark model to obtain OZI relations for neutral
mesons\REF{\ALS}{G. Alexander, H.J. Lipkin and F. Scheck,
Phys. Rev. Lett. 17 (1966) 412}$[{\ALS}]$
was the selection rule forbidding reactions like
$$ \sigma (\pi^-p  \rightarrow N\phi)=0 \eqno(YY4.1a) $$
and its SU(3) rotation predicting the equality
$$ \sigma (K^-p  \rightarrow \Lambda \omega) =
\sigma (K^-p  \rightarrow \Lambda \rho) \eqno(YY4.1b) $$
The $\rho^o$ and $\omega$ mesons are produced in the reactions (YY4.1b) only
via their $u \bar u$ component and thus are produced equally.

An outstanding failure of a quantitative prediction of
an OZI-forbidden process was the experimental discovery of the $J/\psi$
by pure accident while no theorist had predicted the narrow width
nor directed experimenters to look for these enormous signals.
The big charm-search review paper by Gaillard, Lee and Rosner
\REF{\GLR}{M K. Gaillard, B.W. Lee and J.L. Rosner, Rev. Mod. Phys.
47 (1975) 277}$[{\GLR}]$ predicted the vector charmonium state, overestimated
its width by a factor of 30, and did not point out the striking signal of a
very narrow resonance. The very narrow width
caused considerable confusion after the discovery of the $J/\psi$
and was used as evidence against the charmonium interpretation. Feynman
insisted that this ``crazy Zweig rule" could not give such a large suppression,
because it was violated by two-step strong interaction processes where
each step was allowed and perturbation theory was certainly not valid.
There must be some new symmetry principle with a new conserved quantum
number.

This failure to understand the OZI rule led to overestimating the width by a
factor of 30.  The experimental $\phi \rarrow \rho \pi$ width was used as
input$[{\GLR}]$ and threshold effects were disregarded. But the
$\phi \rarrow \rho \pi$ decay is dominated by the two-step transition
$\phi \rarrow K \bar K \rarrow \rho \pi$ for which
the OZI-allowed $K{\overline K}$ channel is open.
The use of the experimental $\phi \rarrow \rho \pi$
width as input can give only an upper bound for the width of the
$J/\psi$ decay where no OZI-allowed channel is open and
the $D{\overline D}$ channel analogous to $K{\overline K}$ in
$\phi \rarrow \rho \pi$ is closed.
The distinction between open on-shell and closed
off-shell intermediate states is now known to be
significant because the physically observable transitions to open
on-shell channels are related by unitarity to the OZI-forbidden
processes
\REF{\HJLOZI}{Harry J. Lipkin,
in {\it New Fields in Hadronic Physics},
Proceedings
of the Eleventh Rencontre de Moriond, Flaine - Haute-Savoie, France,
1976 edited by J. Tran Thanh Van (Rencontre de Moriond, Laboratoire de
Physique Theorique et Particules Elementaires, Universite de Paris-Sud,
Orsay, France, (1976) Vol. I. p. 169,
and {\it Phys. Lett.} {\bf B179} (1986) 278}
$[{\HJLOZI}]$
and because the amplitudes via on-shell
intermediate states cannot be canceled by off-shell contributions.
But there still is no real answer to Feynman's argument against the
narrowness of the $J/\psi$. Hand-waving arguments suggest that
second order processes are cancelled by contributions from different
intermediate states. But there is still no rigorous QCD argument
supported by calculations.

The GLR paper $[{\GLR}]$ contains a note attributed to me, suggesting $e^+-e^-$
as the best place to look for charm,
since the charge +2/3 gave a much larger relative cross section.
The most striking signal would be a large increase in the number of
strange particles, since charm would decay to strangeness. Half of the
hadronic events above charm threshold would contain strange particles.
My argument was correct but the signal was not seen.
At the 1975 Lepton-Photon Conference Haim Harari resolved the paradox, by
noting that the excess of strange particles was not
observed because of the unexpected appearance near charm threshold of the tau
lepton. The nonstrange hadrons from $\tau$ events compensated for the
strangeness excess from charm. At that time the existence of the $\tau$ as well
as the identification of the $J/\psi$ as charmonium were still controversial.
\medskip

\centerline{\bf V. ABSENCE OF FREE QUARKS AND FRACTIONAL CHARGES}
\medskip

Much of the resistance of the particle physics establishment to the quark
model was based upon their fractional charge and upon the failure of
experimenters to find free quarks. Both points are red herrings.
\medskip

\centerline{\bf 5.1 Why There Are No Free Quarks}

Why should anyone expect to find free quarks? A so called ``free electron" is
a very complicated object containing a cloud of virtual photons and $e^+-e^-$
pairs. The hydrogen atom is much more than a point electron and a point
proton,.
The other constituents are observed in Lamb shift and other experiments.
Theorists describe this complicated structure only by using infinite
renormalizing constants.
Pulling the hydrogen atom apart into an electron and a proton, each containing
its own infinite cloud of junk, was possible because the vacuum polarization
between the electron and proton was small when they were separated. The energy
required to excite and ionize the hydrogen atom was less than the rest mass
of an electron-positron pair by a factor of order $10^5$.

But suppose the
excitation energy of the first excited state of the hydrogen atom was more
than double the mass of positronium. The excited states would decay almost
immediately by emitting positronia and isolated electrons would not have
been discovered. Hitting the electron with a photon having enough energy to
move it far away from the proton would polarize the vacuum and create a string
of electron-positron pairs, which would quickly recombine into neutral
positronia. Atomic collisions could well produce ``electron jets" of neutral
atoms and positronia and no free electrons).
Free constituents would not be easily found for hadrons whose spectrum
indicated a structure with the energy of the first excited state already
greater than twice the pion mass. The energy required to move these
constituents from their lowest orbit into the first excited orbit was already
greater than double the rest mass of the lowest bound state. Thus pumping
energy into the proton would simply create pions and other bound states. The
forces and vacuum polarization created by trying to remove a quark from a
proton were much too great to allow the quark to be removed like the electron
from a hydrogen atom.

Already in the late 1960's the hadron spectrum suggested that hitting a quark
produced a string of pairs, and that the excitation spectrum looked like the
spectrum of a string $\({\Venez}\)$.
One does not have to invent fancy names like confinement
and chromodynamics to understand this simple physics.
But the establishment refused to budge from its reactionary position.
The party line that nothing was more elementary than neutrons and protons
was sacrosanct and heretics were ridiculed.
\medskip

\centerline{\bf 5.2 Who Needs Integrally Charged Quarks}

The prejudice against fractional charge led to a number of proposals of
models with integrally charged quarks and a series of useless
proposals for experiments to measure the quark charge.
The basic fallacy in the arguments for and against integral charge is
seen by noting that the electromagnetic current must have a color octet
component in all models with integrally charged quarks, and that all
matrix elements of color octet operators vanish between color singlet
states. Thus all experiments involving only color singlet hadrons can
measure only the color singlet component of the quark charge and will
give the fractional charge\REF{\HJLFRAC}{Harry J. Lipkin,
Phys. Rev. Lett. 28 (1972) 63; Phys. Lett. {B85} (1979) 236;
Nuc. Phys. B135 (1979) 104} $[{\HJLFRAC}]$.

If quarks really have integral charge but color octet hadrons exist
only at the Planck mass, there is no way to observe the integral charge
at reasonable energies and therefore no way to kill the integrally charged
models. Looking for evidence for integrally charged
quarks is useless far below the threshold for producing color
octet states. The only sensible answer to the proposal that quarks might
have integral charge is ``Who needs them"? Why bother shooting
down such models? One can paraphrase Pauli's remark about hidden
variables: ``Integrally charged quark models are like mosquitoes - the
more you kill, the more there are."

\refout

\end